\documentclass[pra,aps,longbibliography,twocolumn,superscriptaddress,floatfix]{revtex4-1}
\usepackage{amsmath,amssymb,graphicx}
\usepackage[usenames,dvipsnames]{color}
\usepackage{verbatim}
\usepackage{setspace}
\usepackage{dcolumn}
\usepackage{bm}
\usepackage{booktabs}
\usepackage{capt-of}
\usepackage{siunitx}
\usepackage{mathrsfs}
\usepackage[T1]{fontenc}
\usepackage[bookmarks=false,linkcolor=blue,urlcolor=blue,colorlinks,citecolor=blue]{hyperref}

\graphicspath{{figures/}}
\begin{document}

\title{LLM-Driven Cross-Paradigm Design for Quantum Optimal Control}

\author{Yu-Qin Chen}
\email{yqchen@gscaep.ac.cn}
\affiliation{Graduate School of China Academy of Engineering Physics, Beijing 100193, China}

\author{Shi-Xin Zhang}
\email{shixinzhang@iphy.ac.cn}
\affiliation{Institute of Physics, Chinese Academy of Sciences, Beijing 100190, China}

\date{Draft: 18 June 2026}

\begin{abstract}
Quantum optimal control (QOC) underpins adiabatic quantum computation, quantum annealing, and quantum state engineering, yet practical deployment is fundamentally bottlenecked by strict hardware constraints and substantial expert effort required to design protocols for each problem instance. To overcome this, we introduce QOC-Workbench, an auditable, large language model (LLM)-driven workflow that acts as an automated quantum co-scientist for cross-paradigm protocol design.
Going beyond traditional numerical optimizers that merely tune parameters within a fixed formula, the LLM autonomously parses physics literature, proposes structural hypotheses, and writes code to validate them by direct simulation. This workflow supports cross-paradigm design by accumulating control motifs across tasks. We demonstrate this approach across three distinct settings: Case 1, Rydberg-atom maximum-independent-set arrays; Case 2, interacting XXZ spin chains; and Case 3, random transverse-field Ising models. In Cases 1 and 2, the workflow autonomously discovers hardware-compliant auxiliary controls, target catalysts, and schedule deformations that outperform literature baselines. In Case 3, it addresses the computational bottleneck of variational counterdiabatic driving by escalating from per-instance optimization to an amortized graph-neural-network generator, successfully transferring learned coefficient paths to larger unseen systems. By actively bridging the gap between theoretical algorithms and experimental restrictions across distinct control paradigms and Hamiltonian families, QOC-Workbench establishes a continuously evolving, cross-paradigm methodology for autonomous quantum control.
\end{abstract}

\maketitle

\section*{Introduction}
Rapid progress across quantum hardware platforms including superconducting processors~\cite{kim2023evidence,google2023suppressing}, trapped-ion simulators~\cite{monroe2021programmable}, and neutral-atom arrays~\cite{saffman2016quantum,henriet2020quantum,browaeys2020manybody,Ebadi2022science,bluvstein2024logical,zhang2026acqc} has brought practical quantum computation and simulation closer to reality~\cite{preskill2018nisq}. However, extracting maximal performance from these finite coherence time, noise-limited devices remains a formidable challenge. Quantum optimal control (QOC) strategies, such as adiabatic quantum computation (AQC), quantum annealing, and broader state engineering~\cite{born1928adiabatic,kato1950adiabatic,farhi2000quantum,farhi2001quantum,kadowaki1998quantum,aharonov2008adiabatic,albash2018adiabatic,hauke2020perspectives}, attempt to bridge this gap by encoding target tasks into continuous-time Hamiltonian evolution. While advanced techniques like approximate counterdiabatic (CD) driving~\cite{demirplak2003adiabatic,berry2009transitionless,delcampo2013shortcuts,kolodrubetz2017geometry,sels2017minimizing,guery2019shortcuts,hegade2022digitized}, schedule optimization~\cite{roland2002quantum,nmi2020rl,finzgar2024designing,perdomo2018opportunities}, and path catalysts~\cite{Albash2019cat, designing2020,xxz2026local} can effectively mitigate diabatic errors, they share a critical bottleneck: establishing a successful protocol requires exhaustive, case-by-case expert design. Because each platform possesses distinct control primitives, pulse bounds, and spectral properties, every proposed strategy must be manually tailored to both the specific target problem and the strict native constraints imposed by the hardware. This reliance on bespoke, hand-crafted protocol design---a trial-and-error process that traditionally consumes days or weeks of human expert time per specific instance---has become a major roadblock in scaling quantum utility.

Historically, quantum control protocol development has been bounded by what we term \emph{interpolation}: the refinement of parameters within a single predefined strategy whose functional form is fixed before optimization begins. For instance, local adiabatic schedules and modern reinforcement-learning or Bayesian-optimization approaches can optimize the coefficients of a chosen nonlinear schedule function~\cite{roland2002quantum,nmi2020rl,finzgar2024designing}, whereas variational CD methods tune auxiliary generators along a specified Hamiltonian path using a predetermined operator ansatz~\cite{sels2017minimizing,weightedcd2026local}. Even when attempting to combine multiple approaches, the search space is still a closed list of candidate strategies and their combinations~\cite{designing2020,xxz2026local}. 

Crucially, such manual workflows cannot autonomously propose new functional forms, introduce previously unconsidered structural modifications to the evolution path, or discover qualitatively different control families beyond what the human designer has enumerated. The design space is closed at the outset: the optimizer refines parameters but cannot reshape the space itself. This rigid boundary between \emph{what is searched} and \emph{what can be proposed} is the fundamental limitation we seek to overcome. Large language models (LLMs) offer a qualitatively different capability: they can parse physics literature, propose entirely new schedule functional forms or auxiliary Hamiltonian structures, implement them as executable code, evaluate them by direct simulation, and feed the outcome back into an expanding memory~\cite{brown2020language,ouyang2022training,boiko2023autonomous,aygun2026era,gottweis2026coscientist,ghareeb2026robin,lu2026automation}. This makes it possible to continuously enlarge the set of strategies under consideration rather than merely optimizing within a fixed one.

\begin{figure*}[t]
  \centering
  \includegraphics[width=0.94\textwidth]{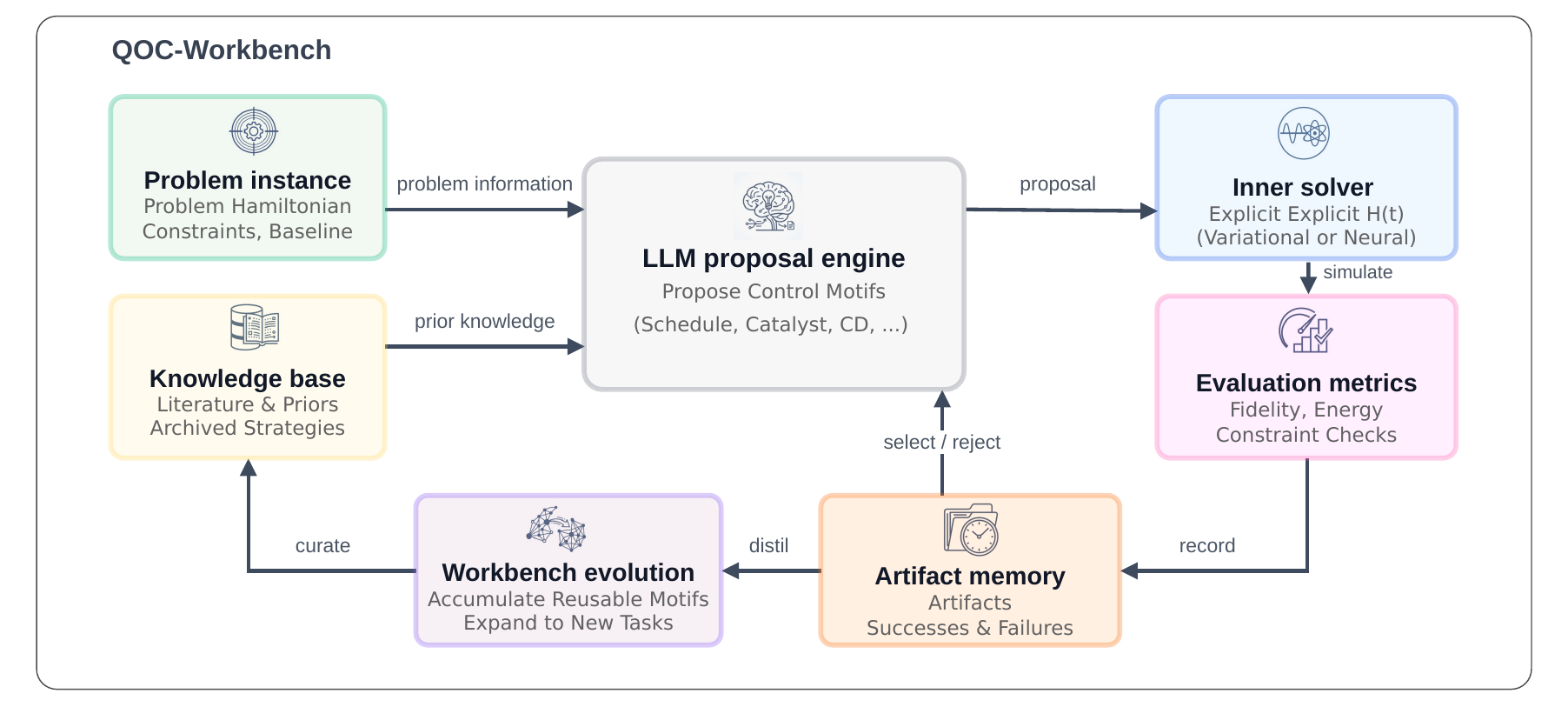}
  \caption{\textbf{LLM-driven cross-paradigm design for quantum control protocols.} The workflow couples Hamiltonian instances, an LLM proposal engine, inner solvers or neural generators, direct quantum simulation and artifact memory. The aim is to make the generation--evaluation--selection loop auditable and transferable across Hamiltonian families.}
  \label{fig:framework}
\end{figure*}

Building on this observation, we develop \emph{QOC-Workbench}, an auditable, LLM-driven platform that acts as an automated quantum co-scientist for cross-paradigm control design. Rather than functioning as a black-box optimizer, the Workbench is designed to emulate the scientific discovery loop of a human physicist. Starting from a structured knowledge base, the Workbench reads established literature to reproduce baselines, hypothesizes physically constrained structural refinements, conducts experiments by executing direct numerical simulations, and accumulates both successes and failures into a persistent memory. Because every outcome is written back into the knowledge base, the workflow functions as a continuously evolving design workbench rather than a one-off optimization script. Over successive tasks, the Workbench accumulates a growing library of reusable control motifs that can be retrieved and adapted to new problems. These motifs include empirical relaxations of analytic CD terms, nonlinear schedule deformations, and amortized neural generators. By delegating the empirical search to the LLM while humans define the problem instances and hardware boundaries, QOC-Workbench establishes a continuously evolving methodology where every discovered protocol remains physically grounded and fully auditable.

We demonstrate this workflow in three case studies that progressively escalate the scope of autonomous design. In hardware-native Rydberg-atom quantum annealing for maximum-independent-set optimization (Case 1), the Workbench starts from a reproduced literature baseline and systematically refines waveform-level controls under strict analog hardware constraints, discovering pulse shapes that consistently outperform the original protocol. In the interacting XXZ spin-chain benchmark (Case 2), the Workbench moves beyond waveform tuning to structural modifications of the many-body evolution path by introducing auxiliary Hamiltonian terms, retuning approximate CD corrections, and reshaping the interpolation schedule. In random transverse-field Ising models (Case 3), the Workbench addresses the computational scalability of per-instance protocol design by converting an analytic control method into a learned, graph-conditioned neural generator that transfers from small to larger system sizes without retraining. Together, these cases validate the core premise of QOC-Workbench.

\section*{Results}

\subsection*{The quantum optimal control problem}

The central task addressed in this work is the following: given a 
target Hamiltonian $H_f$ whose ground state $|\psi_{\rm gs}\rangle$ 
encodes the solution to a computational or physical problem, 
design a time-dependent Hamiltonian $H(t)$ for $t \in [0, T]$ 
such that a system initialized in an easily prepared state 
$|\psi(0)\rangle$ evolves under the Schrödinger equation (setting $\hbar=1$),
\begin{align}
  i \frac{d}{dt}|\psi(t)\rangle = H(t)\,|\psi(t)\rangle,
  \label{eq:schrodinger}
\end{align}
to a final state $|\psi(T)\rangle$ with maximal overlap with 
the target ground subspace,
\begin{align}
  \max_{\{u_k(t)\}} \; F = \langle\psi(T)|\Pi_0^f|\psi(T)\rangle,
  \label{eq:fidelity-objective}
\end{align}
where $\Pi_0^f$ projects onto the ground subspace of $H_f$. 
The Hamiltonian $H(t)$ is typically parameterized by a set of 
time-dependent control functions $\{u_k(t)\}$ subject to 
platform-specific constraints such as bounded amplitudes, 
restricted control channels and fixed interaction terms.

Adiabatic quantum computation approaches this problem by 
interpolating between an initial Hamiltonian $H_i$ with a 
known ground state and the target $H_f$,
\begin{align}
  H_{\rm AQC}(t) = [1 - \lambda(t)]\,H_i + \lambda(t)\,H_f 
  + H_{\rm aux}[\lambda(t)],
  \label{eq:aqc-path}
\end{align}
where the schedule $\lambda: [0,T] \to [0,1]$ and optional 
auxiliary terms $H_{\rm aux}$ constitute the design degrees 
of freedom. The adiabatic theorem guarantees that $|\psi(T)\rangle 
\to |\psi_{\rm gs}\rangle$ in the limit $T \to \infty$, but 
finite coherence times require short $T$, and diabatic transitions 
through small spectral gaps can strongly affect fidelity.

This formulation makes explicit why the design problem is 
fundamentally more challenging than standard parametric 
optimization. The search space is not a finite-dimensional 
parameter vector but a space of \emph{control functions} 
$\{u_k(t)\}_{k=1}^K$ together with structural choices: which 
auxiliary terms to include, how to parameterize the schedule, 
whether to add CD corrections, and how to 
constrain the control channels to respect hardware limitations. 
No single gradient-based optimizer or hardcoded search algorithm 
can efficiently navigate this hybrid continuous--discrete design 
space across different Hamiltonian families. It is precisely 
this challenge—searching over functional forms, structural 
choices and physical paradigms simultaneously—that motivates 
the LLM-driven workflow described below.

To validate this framework, the subsequent sections demonstrate progressively deeper levels of autonomous design: from waveform-level control refinement within a fixed physical architecture (Case 1), to Hamiltonian-path engineering (Case 2), and finally to amortized protocol generation where transferable neural policies are trained for entire families of Hamiltonians (Case 3).

\begin{figure*}[t]
  \centering
  \includegraphics[width=0.95\textwidth]{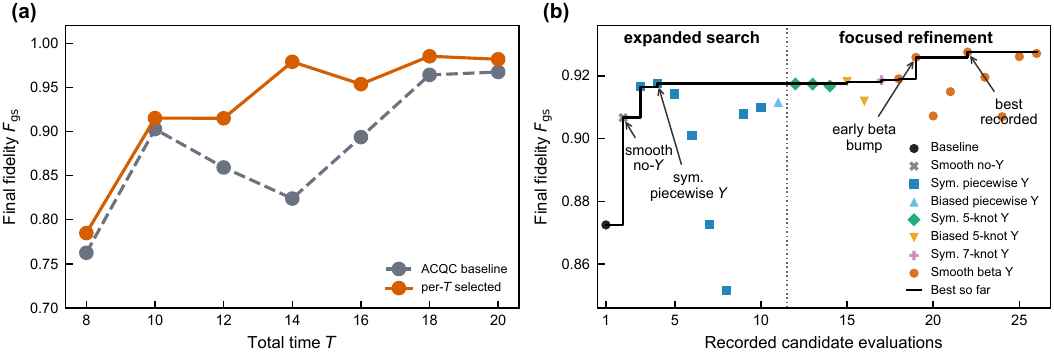}
  \caption{\textbf{Rydberg MIS control-pulse design and exploration history.} \textbf{a}, Fixed C6 instance evaluated at seven total times. The gray dashed curve shows the final fidelity of the reproduced smooth ACQC baseline as a function of $T$. The orange curve shows the final fidelity of the best candidate selected independently at each $T$ from 62 candidate slots evaluated separately for each total time. Averaging the seven plotted values gives $0.882$ for the baseline and $0.931$ for the selected protocols. \textbf{b}, Ten-site C10 search trace at fixed $T=30$. Points show hardware-compatible candidates proposed by QOC-Workbench, recorded in chronological order. The black staircase marks the best final fidelity found so far, and the vertical dotted line separates the expanded search from the focused refinement. The best early-biased smooth beta-bump pulse reaches final fidelity $0.9275$, compared with $0.8725$ for ACQC. Protocol abbreviations used as point labels are detailed in the Supplementary Information.}
  \label{fig:rydberg}
\end{figure*}

\subsection*{An auditable LLM-driven search loop}
At the core of QOC-Workbench is an iterative search loop for protocol design, summarized in Fig.~\ref{fig:framework}. The platform functions as a continuously evolving design workbench. When a new task is introduced, the workflow systematically retrieves relevant priors, proposes feasible modifications, evaluates them via simulation, and stores the outcome back into the Workbench. Every accepted claim is backed by clear data provenance, and every failure is recycled into useful experience.

\textbf{Knowledge base.} QOC-Workbench contains a curated knowledge layer that precedes any new design round. This layer stores research papers, extracted Hamiltonian specifications, reported baselines and other design-relevant information. The purpose is not merely archiving documents, but converting heterogeneous materials into searchable design ingredients that the LLM can cite, compare and translate into executable protocols.

\textbf{Problem instance.} A new search begins by specifying a concrete task: the explicit form of the target Hamiltonian, constraints imposed by the experimental platform or numerical setting, and an optional reference path or baseline schedule. This explicit task specification sets the physical search space before the LLM proposes protocol modifications.
This task-formalization step also optionally includes choosing an operator representation, an initial-state preparation rule, admissible or explicitly forbidden control channels and the task-specific evaluation metric, which together define the executable framework for subsequent search rounds.

\textbf{LLM proposal engine.} The experiments reported here used OpenAI Codex with GPT-5.5 as the coding and reasoning system.
Notably, the proposal engine does not simply tune given parameters. It formulates concrete physical hypotheses and constructs the implementation required to test them against exact quantum simulation.
These physically motivated candidate moves include modifying the evolution schedule, adding or changing catalysts and auxiliary terms, testing different CD constructions, or introducing entirely new control families such as neural generators. Importantly, the design space is not fixed in advance as a closed list of human-written choices. New candidates can arise from previous attempts, from methods learned from the literature, from analogies across systems, or from exploratory ideas that go beyond conventional quantum control design. 
Cross-paradigm design emerges naturally from this workflow with reusable control motifs across different physical contexts.

\textbf{Inner solver or neural generator.} After a design proposal is made, it must be realized as an explicit time-dependent Hamiltonian. We use the term \emph{solver} for modules that compute control coefficients by solving a variational or algebraic problem, and \emph{generator} for modules that directly produce control paths from analytic formulas, piecewise parameters or trained neural networks. In other words, the LLM suggests a control family, but the solver or generator produces the numerical Hamiltonian actually used in time evolution. Once a simulation backend has been defined, it is treated as protected infrastructure: candidate improvements must arise from admissible protocol modifications rather than from changing the simulator, target Hamiltonian or evaluation rule, to ensure that improvements reflect genuine protocol quality rather than artifacts of modified evaluation criteria. Implementation details are described in Methods.

\textbf{Evaluation and memory.} Each candidate is selected or rejected by simulation metrics matched to the task, including final ground-subspace fidelity, normalized energy, adiabatic-following metrics, coefficient prediction error and hardware-constraint checks. Results are stored in timestamped artifacts containing executed code, Hamiltonian descriptions, parameter files, and outputs. Memory also archives unsuccessful trials, which act as boundary conditions to narrow the future search space and prevent the workflow from revisiting dead ends. At the next round, selected tables and metrics are retrieved and supplied as compact context, turning the search into a gradient-free, feedback-driven evolution over protocol families. This iterative workflow terminates based on predefined stopping criteria, such as exceeding a target performance threshold or exhausting a search budget.

QOC-Workbench is designed to grow as new quantum optimal control methods, hardware platforms and problem Hamiltonians are introduced. A method can first enter as a literature note, then become an executable solver, a learned generator or a reusable search primitive; a failed idea remains as a boundary condition for later reasoning. The long-term goal is that each new task can be approached by the same cycle: formalize, retrieve, generate, simulate, record and transfer.

\begin{figure*}[t]
  \centering
  \includegraphics[width=0.95\textwidth]{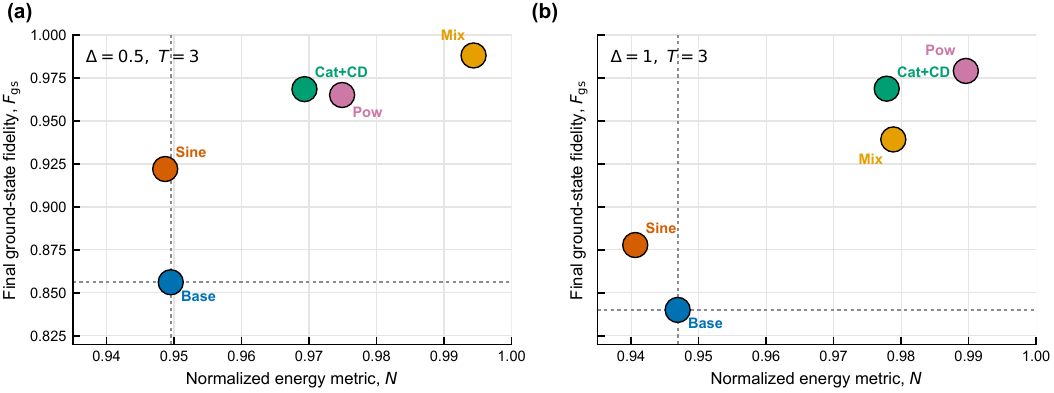}
  \caption{\textbf{XXZ catalyst and schedule design.} Both panels use an eight-site periodic XXZ spin chain evolved for total time $T=3$. \textbf{a}, Results for anisotropy $\Delta=0.5$. Points compare the reproduced baseline with representative improved candidates from four LLM-designed families in the plane of normalized final-energy metric $N$ and final ground-state fidelity $F_{\rm gs}$. \textbf{b}, Results for anisotropy $\Delta=1$ after replacing endpoint-nonsmooth schedule variants by endpoint-clean alternatives, shown with the same axes and protocol colors. Dashed guide lines mark the reproduced baseline in each panel. Protocol abbreviations used as point labels are detailed in the Supplementary Information.}
  \label{fig:xxz}
\end{figure*}

\subsection*{Case 1: Hardware-compliant pulse design on Rydberg arrays}
The first case study evaluates the workflow on a hardware-native quantum annealing task: solving the maximum-independent-set (MIS) problem, the task of finding the largest subset of mutually non-adjacent vertices in a graph, on analog Rydberg-atom arrays~\cite{Ebadi2022science}:
\begin{equation}
H_{\mathrm{Ryd}}(t)={}\frac{\Omega(t)}{2}\sum_i X_i
 - \Delta(t)\sum_i n_i
+ V\sum_{(i,j)\in E}n_in_j.
\label{eq:rydberg}
\end{equation}
Here $n_i=(I-Z_i)/2$ is the excitation operator, $\Omega(t)$ the global Rabi drive, $\Delta(t)$ the detuning, and $V$ the fixed interaction penalty. The target MIS embeds naturally into the Rydberg Hamiltonian due to the geometric blockade effect: the strong repulsive interaction $V$ prohibits connected atoms from being simultaneously excited. The Workbench's task is to optimize the continuous-time evolution to maximize target ground-state fidelity within strict analog hardware constraints: $V$ is fixed, $\Omega(t)\geq 0$, and auxiliary controls are restricted to global one-body quadratures $f_Y(t)\sum_iY_i$.

The Workbench began by reproducing the analytic CD quantum control (ACQC) protocol~\cite{zhang2026acqc}. However, the Workbench recognized that this analytic coefficient is derived solely from the non-interacting sector, rendering it too rigid for the full many-body system. It autonomously evolved a sequence of increasingly flexible design families that act as empirical relaxations of the analytic CD-inspired waveform. This progression included rescaling the analytic pulse, proposing variationally optimized multi-knot waveforms, and jointly adjusting the main analog schedules via piecewise controls.

Evaluated on a six-site cycle (C6) across various total evolution times (Fig.~\ref{fig:rydberg}a), the Workbench-designed protocols consistently outperformed the ACQC baseline, improving average fidelity from $0.882$ to $0.931$. The search revealed that the optimal strategy is timescale-dependent: short times favor rescaled analytic pulses, while intermediate times require joint piecewise schedule deformations and multi-knot pulses (see Supplementary Fig.~\ref{fig:app-c6-controls} for waveform visualizations). On a larger ten-site instance (C10), the search history reveals how Workbench explored increasingly flexible control families, progressing from rescaled analytic pulses through piecewise global-$Y$ waveforms to a smooth beta-bump envelope, ultimately converging on the best protocol through refinement (Fig.~\ref{fig:rydberg}b). It discovered that an early-biased smooth beta-bump pulse yielded the highest fidelity, demonstrating its ability to abstract a control motif and reshape its functional envelope under fully interacting many-body dynamics. It is worth noting that spectral analysis (see Supplementary Information) reveals the physical mechanism behind this success: rather than simply maximizing the minimum spectral gap as conventionally assumed, the optimized protocol reshapes the low-energy instantaneous spectrum and shifts the location of the bulk-gap bottleneck relative to the ACQC baseline to achieve better performance.

\begin{figure*}[t]
  \centering
  \includegraphics[width=0.97\textwidth]{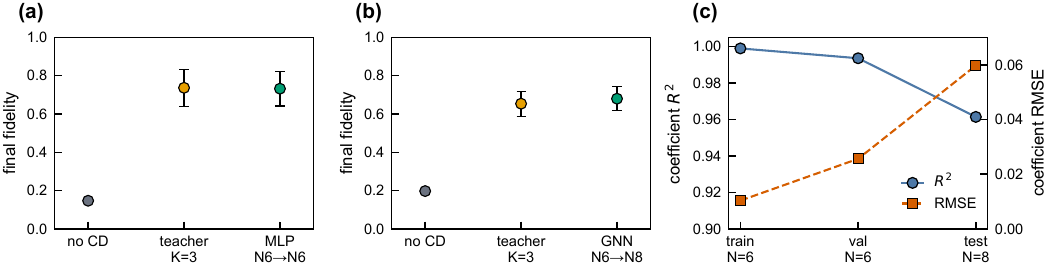}
  \caption{\textbf{Amortized weighted-CD generation for random TFIM instances.} \textbf{a}, Fixed-size generalization at $T=1.5$ for no CD, exact weighted-CD teacher and the node-wise MLP generator on 12 unseen $2\times3$ ($N=6$) test instances. Points show mean final fidelity and error bars denote one standard deviation over test instances. \textbf{b}, Cross-size transfer at $T=2.5$ for no CD, exact weighted-CD teacher and the GNN generator on six unseen $2\times4$ ($N=8$) test instances, with the same mean and standard-deviation notation. Here $K=3$ denotes the hyperparameter for the weighted variational-CD teacher loss, which biases the fit toward the low-energy subspace; details are given in the Supplementary Information. \textbf{c}, Coefficient prediction quality for the GNN, evaluated on the training split, held-out validation split and zero-shot test split. Blue circles show coefficient $R^2$ on the left axis, and orange squares show RMSE on the right axis.}
  \label{fig:tfim}
\end{figure*}

\subsection*{Case 2: Catalyst and schedule co-design on XXZ spin chains}
Building on the schedule-reshaping and empirical CD relaxation motifs, we next evaluate the workflow's application to Hamiltonian-path engineering in an interacting XXZ spin chain, a canonical model with nontrivial spectral structure~\cite{albash2018adiabatic,xxz2026local}. The target Hamiltonian on a ring is:
\begin{equation}
H_{\rm XXZ}=\frac{1}{4}\sum_{j=1}^{8}\bigl(X_jX_{j+1}+Y_jY_{j+1}+\Delta Z_jZ_{j+1}\bigr).
\label{eq:xxz}
\end{equation}

The Workbench first surveyed the literature~\cite{xxz2026local} and reproduced a strong reference baseline combining optimized initialization, auxiliary Hamiltonians, and approximate CD. From this reproduced baseline, the Workbench moved beyond simple parameter tuning to jointly optimize the Hamiltonian path, the interpolation schedule, and the CD corrections. It introduced an endpoint-vanishing target catalyst, retuned the commutator-CD coefficients, and tested various nonlinear schedules, including mixed-power, pure-power, and sine deformations (see Supplementary Fig.~\ref{fig:app-xxz-schedule} for schedule visualizations). These nonlinear schedules effectively redistribute the evolution time along the interpolation path; simulation results demonstrate that this temporal redistribution yields additional performance gains in finite-time evolution compared to the baseline. Furthermore, for schedule variants sensitive to endpoint behavior, the Workbench autonomously applied windowed CD corrections to guarantee that auxiliary terms cleanly vanish at the boundaries.

Tested at anisotropy values $\Delta=0.5$ and $\Delta=1$ (Fig.~\ref{fig:xxz}), the most successful protocols combined a target catalyst with retuned CD and nonlinear schedule deformations. This co-design, utilizing mixed-power and pure-power schedules, further outperformed the catalyst/CD baseline across the reported metrics, with the best candidate depending on the specific metric and Hamiltonian parameter. For example, at $\Delta=0.5$ a mixed-power schedule performs best globally, whereas at $\Delta=1$ a pure power schedule is superior for final fidelity and energy while the mixed schedule optimizes average adiabatic tracking. Crucially, the Workbench found that performance gains did not come from parameter retuning alone; the best protocols required structural optimization of the Hamiltonian and the interpolation path.

\subsection*{Case 3: Amortized path generation on transverse-field Ising models}
In the previous cases, the workflow operated by optimizing parameters for each specific Hamiltonian instance or target time. Case 3 addresses a different bottleneck: the cost of repeating this parameter optimization for every new instance. Here the Workbench transitions from single-instance protocol search to amortized generation across random transverse-field Ising models (TFIM) on rectangular lattices:
\begin{align}
H(\lambda) ={}&(1-\lambda)\sum_i X_i
\nonumber\\
&+ \lambda\left(\sum_i h_i Z_i
- \sum_{\langle i,j\rangle}J_{ij}Z_iZ_j\right),
\label{eq:tfim}
\end{align}
where $h_i$ and $J_{ij}$ are random local fields and couplings.

The Workbench identified the weighted-action variational-CD framework~\cite{weightedcd2026local} as a powerful analytic prior. This method computes local CD coefficients $\boldsymbol{\alpha}$ by minimizing a polynomially weighted variational action to bias the correction toward the low-energy subspace. However, the Workbench recognized its computational bottleneck: the local optimization
must be repeatedly solved for every graph instance and interpolation time step.

To resolve this, the Workbench proposed a structural paradigm shift: rather than continuously optimizing individual control parameters, it designed an amortized neural generator. It first trained a fixed-size node-wise multilayer perceptron (MLP) as an in-distribution proof of concept, successfully matching the exact teacher's fidelity distribution on small held-out test instances (Fig.~\ref{fig:tfim}a). To enable cross-size generalization, the Workbench deduced that because the target CD operator is constrained to a sum of local one-body terms, a message-passing architecture is well matched to this physical inductive bias. Consequently, it upgraded the architecture to a graph neural network (GNN)~\cite{kipf2017semi,velickovic2018graph,battaglia2018relational}. Trained exclusively on small instances, the Workbench deployed the GNN zero-shot to larger, unseen lattices (Fig.~\ref{fig:tfim}b,c). Remarkably, the GNN maintained high coefficient prediction accuracy and achieved a mean final fidelity comparable to the exact weighted-CD teacher, completely bypassing the expensive per-instance optimization. This demonstrates the Workbench's capability to abstract physical constraints and engineer transferable computational workflows (see Supplementary Information for more details).

\section*{Discussion}
The demonstrations above establish that quantum optimal control can be successfully automated beyond single-paradigm parameter tuning or one-shot analytic derivation. None of the discovered strategies was hard-coded or explicitly prompted; instead, they emerged from the Workbench's accumulated knowledge, prior experience, and iterative trials.

Cross-paradigm design refers to recognizing abstract control motifs in one context and re-implementing them using the specific control language of new quantum systems. The three case studies instantiate this principle concretely: the workflow crosses the boundary between strict analytic derivations and numerical variational waveforms (Case 1), fuses Hamiltonian structure and schedule paradigms into co-designed protocols (Case 2), and escalates from per-instance optimization to amortized neural path generation (Case 3). Importantly, automating these empirical, scorable search tasks does not replace human expertise. The human scientist remains responsible for reasoning about underlying mechanisms, defining physical constraints, and interpreting the physics behind machine-generated motifs~\cite{aygun2026era,gottweis2026coscientist,ghareeb2026robin,lu2026automation}.

While the current demonstrations provide foundational evidence for LLM-driven cross-paradigm design, several natural avenues remain for future exploration. First, because QOC-Workbench treats hardware constraints as foundational boundaries, extending the framework to incorporate open-system effects, hardware-calibration data, and direct experimental feedback is a natural next step. Second, the design space can be broadened beyond temporal pulse shaping to include hardware mapping and layout co-design, such as jointly optimizing the spatial embedding of a problem graph alongside its temporal evolution schedule. Finally, the underlying paradigm of LLM-driven autonomous discovery extends far beyond quantum optimal control. Because this workflow relies fundamentally on formalizing physical boundaries and evaluating algorithmic hypotheses via simulation, it can be seamlessly generalized to quantum architecture search~\cite{Zhang2020qas, Du2020qas}, quantum many-body physics, and broader scientific computing. Disciplines characterized by large hybrid search spaces and clearly evaluable objectives can benefit from this artifact-driven discovery cycle.

Looking forward, this workflow changes the unit of progress in quantum optimal control protocol design. Instead of reporting a single hand-crafted protocol, each result is linked to executable code, parameter choices, numerical outputs, failed alternatives and reusable memory. This makes it possible to accumulate design knowledge across Hamiltonian families and to ask which control motifs transfer, which fail and which should be promoted into solvers or learned generators. Such artifact-centered accumulation is essential if machine-assisted quantum control is to become a reproducible research methodology rather than a collection of isolated optimized pulses.

\section*{Methods}

\subsection*{Artifact-driven LLM search}
Each search round used a timestamped artifact directory containing the executed code, Hamiltonian description, parameter files, outputs and generated figures. The LLM proposed new feasible modifications to the current protocol, executed simulations through local scripts, inspected metrics and used prior attempts to constrain subsequent proposals. The manuscript uses only physically constrained protocols for the headline comparisons. 

\subsection*{Time-evolution simulation}
In the solver implementation, the dynamics simulation is standardized around TensorCircuit-NG on JAX backend~\cite{zhang2023tensorcircuit,zhang2026tensorcircuitng}. Candidate Hamiltonians are passed to the integrator through matrix-vector-product interface, time evolution is performed by ordinary differential equation integration, and the whole solver is just-in-time-compiled. For each candidate protocol, the solver or generator produced an explicit time-dependent Hamiltonian $H(t)$. The initial state was chosen as a ground state of the initial Hamiltonian, and final fidelity was computed as the squared overlap with the final ground subspace. When the final ground space was degenerate, the fidelity was summed over the degenerate eigenvectors. 

\subsection*{Search and parameter selection}
The search combines heuristics, low-dimensional scans and randomized candidate pools rather than a single global optimizer. In Case 1, the baseline control language was inherited from ACQC: smooth Rabi and detuning schedules were supplemented by a global $Y$-quadrature correction. The scaled-ACQC family multiplied the analytic $Y$ coefficient by a scalar $\alpha$. The piecewise family allowed multi-knot $\Omega(t)$ and $\Delta(t)$ schedules, optionally with a global $Y$ pulse, while enforcing non-negative and bounded Rabi amplitudes, bounded detuning values and a time-independent interaction penalty (see Supplementary Information). In Case 2, the reproduction stage evaluated the Ref.~\cite{xxz2026local} baseline. The subsequent search scanned endpoint-vanishing target catalysts, commutator-CD coefficients, and non-linear schedule deformations (see Supplementary Information). 

\subsection*{Evaluation metrics}

For Case 1 and Case 3, the primary reported metric is the final overlap with the target ground subspace. For Case 2 we report final energy together with a normalized energy metric, the average adiabatic-following metric and final ground-state fidelity. Let $E_0$ be the target ground energy of $H_f$, let $E_{\rm init}=\langle\psi(0)|H_f|\psi(0)\rangle$ be the energy of the initial state measured by the target Hamiltonian and let $E_{\rm final}=\langle\psi(T)|H_f|\psi(T)\rangle$. The normalized energy metric is
\begin{align}
N=\frac{E_{\rm init}-E_{\rm final}}{E_{\rm init}-E_0},
\end{align}
with larger values indicating lower final energy and $N=1$ corresponding to the target ground energy. Thus $E_{\rm final}$ and $N$ quantify the same energy improvement, with $N$ providing a dimensionless comparison across protocols. The average adiabatic-following metric is
\begin{align}
\label{eq:fad}
F_{ad}=\frac{1}{T}\int_0^T \sqrt{\langle\psi(t)|\Pi_0(t)|\psi(t)\rangle}\,dt,
\end{align}
where $\Pi_0(t)$ projects onto the instantaneous ground subspace of the Hamiltonian used for the reference adiabatic path. This metric measures how closely the state follows the instantaneous ground subspace throughout the evolution. The final ground-state fidelity is
\begin{align}
F_{\rm gs}=\langle\psi(T)|\Pi_0^f|\psi(T)\rangle,
\end{align}
where $\Pi_0^f$ projects onto the final target ground subspace. 
Error bars in the plots denote one standard deviation over test instances. Coefficient prediction quality for the neural generator was evaluated against the weighted-CD teacher labels by flattening all coefficients over graph instances, $\lambda$ grid points and sites. For teacher values $y_m$, predictions $\hat y_m$ and $M$ total coefficient samples, the root-mean-square error is
\begin{align}
{\rm RMSE}=\sqrt{\frac{1}{M}\sum_{m=1}^{M}(\hat y_m-y_m)^2},
\end{align}
and the coefficient of determination is
\begin{align}
R^2=1-
\frac{\sum_{m=1}^{M}(\hat y_m-y_m)^2}
{\sum_{m=1}^{M}(y_m-\bar y)^2},
\qquad
\bar y=\frac{1}{M}\sum_{m=1}^{M}y_m.
\end{align}
Thus RMSE measures the absolute coefficient error in the same units as $\alpha_i(\lambda)$, while $R^2=1$ indicates perfect coefficient prediction and $R^2=0$ corresponds to predicting only the teacher mean.

\subsection*{Data and code availability}
All simulation data, parameter files and generated figures used in this work are available from the corresponding authors upon reasonable request. Source code and artifact directories will be deposited in a public repository upon publication.

\section*{Acknowledgements}
This work was supported by the National Natural Science Foundation of China (No. 12504599 and No. 12574546), Quantum Science and Technology-National Science and Technology Major Project (No. 2024ZD0301700 and No. 2025ZD0300802), Science Challenge Project (No. TZ2025017), and the Chinese Academy of Sciences (No. XDB1680201 and No. YSBR-150).

\bibliographystyle{apsreve}
\bibliography{ref}

\clearpage
\onecolumngrid
\clearpage

\renewcommand{\thefigure}{S\arabic{figure}}
\renewcommand{\theHfigure}{S\arabic{figure}}
\setcounter{figure}{0}
\renewcommand{\theequation}{S\arabic{equation}}
\renewcommand{\theHequation}{S\arabic{equation}}
\setcounter{equation}{0}
\renewcommand{\thesection}{\Roman{section}}

\setcounter{section}{0}
\renewcommand{\thetable}{S\arabic{table}}
\setcounter{table}{0}
\setcounter{secnumdepth}{4}

\section*{Supplementary Information}
\setcounter{section}{0}
\renewcommand{\thesection}{Supplementary Note \arabic{section}}
\section{Rydberg candidate pools and finite-time diagnostics}
\label{app:rydberg-details}

This Supplementary Note collects the supplemental Rydberg results supporting Fig.~\ref{fig:rydberg}. The C6 subsection documents the per-time candidate pool and selected controls, while the C10 subsection summarizes the follow-up global-$Y$ waveform refinements and finite-time diagnostics.

\subsection{Rydberg MIS graph instances}
\label{app:rydberg-instances}
The Rydberg examples use cycle-graph MIS instances. The C6 instance has vertices $i=0,\ldots,5$ and edges $(i,i+1\,\mathrm{mod}\,6)$; the C10 instance has vertices $i=0,\ldots,9$ and edges $(i,i+1\,\mathrm{mod}\,10)$. In both cases the simulated Hamiltonian is
\begin{align}
H(t)={}&\frac{\Omega(t)}{2}\sum_i X_i
 -\Delta(t)\sum_i n_i
  +V\sum_{(i,j)\in E}n_in_j
\nonumber\\
&+f_Y(t)\sum_iY_i,
\end{align}
with $n_i=(I-Z_i)/2$, $\Omega_0=\Delta_0=1$ and fixed blockade penalty $V=10$. Protocols without auxiliary quadrature set $f_Y(t)=0$. At the final time the smooth schedules have the boundary condition $\Omega(T)=0$ and $\Delta(T)=1$, so the problem Hamiltonian is recovered as $H_f=-\sum_i n_i+10\sum_{(i,j)\in E}n_in_j$. Its ground states are the maximum independent sets: ``1010..10''. The smooth reference schedules used in the C10 comparison are
\begin{align}
\Omega(t)={}&\Omega_0\sin^2\!\left[\frac{\pi}{2}\sin\!\left(\frac{\pi t}{T}\right)\right],
\\
\Delta(t)={}&-\Delta_0\cos\!\left(\frac{\pi t}{T}\right),
\end{align}
with $T=30$ for C10. The C6 per-time study uses the same graph Hamiltonian and evaluates total times $T=8,10,\ldots,20$.

\subsection{ACQC-inspired Rydberg $Y$ coefficient}
\label{app:rydberg-acqc}
For the searches, the analytic ACQC-like waveform was computed from the single-atom driving sector of Eq.~\eqref{eq:rydberg}, neglecting the interaction term during the derivation. Up to an irrelevant scalar shift, the noninteracting local Hamiltonian is
\begin{equation}
\label{eq:single_atom}
h_i(t)=\frac{\Omega(t)}{2}X_i+\frac{\Delta(t)}{2}Z_i.
\end{equation}
The corresponding two-level counterdiabatic term is proportional to the adiabatic gauge potential of the field vector $(\Omega,0,\Delta)$, giving
\begin{align}
H^{(0)}_{\rm CD}(t)&=f_Y^{\rm ACQC}(t)\sum_iY_i,
\nonumber\\
f_Y^{\rm ACQC}(t)&=
\frac{\Delta(t)\dot\Omega(t)-\Omega(t)\dot\Delta(t)}
{2[\Omega^2(t)+\Delta^2(t)]},
\label{eq:acqcy}
\end{align}
with the overall sign following the $Y$ and laser-phase convention used in Eq.~\eqref{eq:rydberg}. In hardware language, this term can be implemented by combining the $X$ and $Y$ quadratures into a modified Rabi amplitude and phase. In the present work Eq.~\eqref{eq:acqcy} is used only as an approximate, waveform-level design primitive: it corrects the noninteracting driving sector but does not include the many-body blockade interaction in the gauge-potential calculation.

\subsection{C6 per-time candidate pool and selected controls}
\label{app:c6-pool}

The search for the C6 result in Fig.~\ref{fig:rydberg}a used a fixed six-site cycle graph and evaluated the same 62-slot candidate construction independently at each of the seven total evolution times $T=8,10,\ldots,20$, for a total of $7\times 62=434$ full candidate evaluations. At each $T$, the selected point in Fig.~\ref{fig:rydberg}a is the candidate with the largest final ground-subspace fidelity at that particular total time.

\begin{figure}[t]
  \centering
  \includegraphics[width=0.78\textwidth]{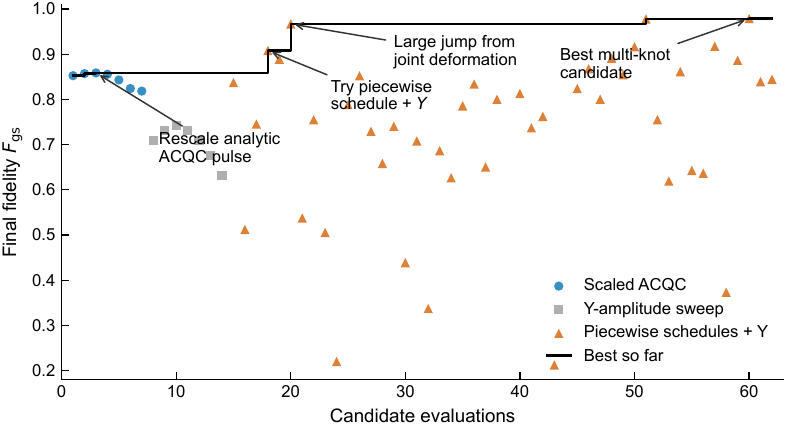}
  \caption{\textbf{C6 candidate-evaluation trace at fixed $T=14$.} Points show the final ground-subspace fidelity of the 62 hardware-compatible candidates in the recorded evaluation order for the $T=14$ C6 pool. Colors and markers denote the protocol family, and the black staircase marks the best fidelity found so far. The selected point in Fig.~\ref{fig:rydberg}a at $T=14$ is the final best multi-knot piecewise schedule-plus-$Y$ candidate in this trace.}
  \label{fig:app-c6-agent-trace}
\end{figure}

The fixed-$T=14$ slice provides a concrete trace of how the selected C6 protocol emerged within one executed candidate pool (Fig.~\ref{fig:app-c6-agent-trace}). The horizontal axis is the recorded candidate-evaluation order. The trace shows that simple ACQC rescaling and smooth $Y$-amplitude sweeps produced only limited gains, whereas the piecewise schedule-plus-$Y$ family contained both low-fidelity failures and the large jumps that ultimately supplied the selected $T=14$ protocol.

\begin{figure}[t]
  \centering
  \includegraphics[width=0.72\textwidth]{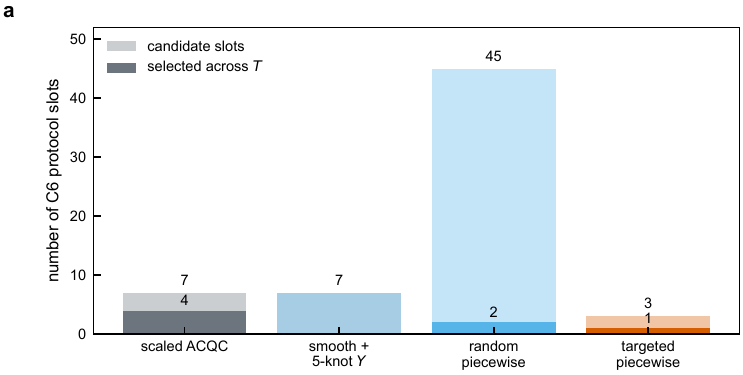}
  \caption{\textbf{C6 candidate families used in the per-time Rydberg search.} Light bars show the number of protocol slots in each family, and filled bars show how often that family was selected as the best protocol across $T=8,10,\ldots,20$. The scaled-ACQC family supplies the best candidate at $T=8,10,12,20$, the random piecewise family at $T=16,18$, and the targeted piecewise family at $T=14$.}
  \label{fig:app-c6-candidate-families}
\end{figure}

Figure~\ref{fig:app-c6-candidate-families} summarizes the 62-slot candidate construction by protocol family. Here a candidate slot denotes one protocol template in the per-time pool; the same family structure and sampling rules were used for each $T$, giving $7\times62=434$ full C6 evaluations across $T=8,10,\ldots,20$. The light bars show how many slots were assigned to each protocol family, whereas the filled bars count how many of the seven per-$T$ winners came from that family.

All multi-knot candidates use normalized-time nodes
\begin{align}
s_k=\{0,1/4,1/2,3/4,1\},
\end{align}
with linear interpolation between adjacent nodes. The scaled-ACQC family keeps the smooth $\Omega(t)$ and $\Delta(t)$ schedules and rescales the analytic ACQC waveform as $f_Y(t)=\alpha f_Y^{\rm ACQC}(t)$, with $\alpha\in\{0.45,0.52,0.60,0.70,0.85,1.00,1.04\}$. The smooth five-knot family also keeps the smooth $\Omega(t)$ and $\Delta(t)$ schedules but replaces the analytic coefficient by
\begin{align}
f_Y(s_k)=\{0,0.4m,m,0.4m,0\},
\end{align}
with $m\in\{-0.12,-0.08,-0.04,0.04,0.08,0.12,0.16\}$.

The random piecewise family contains 45 randomly sampled slots at each total time. For each such slot,
\begin{align}
\Omega(s_k)&=\{0,\omega_1,\omega_2,\omega_3,0\},
\nonumber\\
\Delta(s_k)&=\{-1,\delta_1,\delta_2,\delta_3,1\},
\end{align}
where $\omega_1\sim U(0.35,1.0)$, $\omega_2\sim U(0.55,1.0)$, $\omega_3\sim U(0.20,0.90)$, $\delta_1\sim U(-0.85,-0.25)$, $\delta_2\sim U(-0.15,0.25)$ and $\delta_3\sim U(0.35,1.0)$ in the normalized units of the evaluator. Each random piecewise slot either uses no $Y$ pulse or uses the same symmetric five-knot form $f_Y(s_k)=\{0,0.4m,m,0.4m,0\}$ with $m\sim U(-0.12,0.12)$. The sampling ranges enforce the hardware constraints used in the Rydberg search: non-negative bounded Rabi amplitude, bounded detuning, endpoint-compatible waveforms and fixed interaction strength.

The targeted piecewise family contains three additional non-random templates motivated by earlier search outcomes:
\begin{align}
(\Omega,\Delta,f_Y)(s_k)={}&
(\{0,0.7,1,0.6,0\},\{-1,-0.65,0,0.75,1\},\{0,0.02,0.05,0.02,0\}),
\nonumber\\
&(\{0,0.8,1,0.45,0\},\{-1,-0.75,-0.05,0.65,1\},\{0,0,0,0,0\}),
\nonumber\\
&(\{0,0.55,1,0.75,0\},\{-1,-0.45,0.05,0.85,1\},\{0,-0.02,-0.05,-0.02,0\}).
\end{align}
Thus the implemented per-time pool consists of seven scaled-ACQC slots, seven smooth five-knot $Y$ slots, 45 random piecewise slots and three targeted piecewise slots.

\begin{figure}[t]
  \centering
  \includegraphics[width=0.82\textwidth]{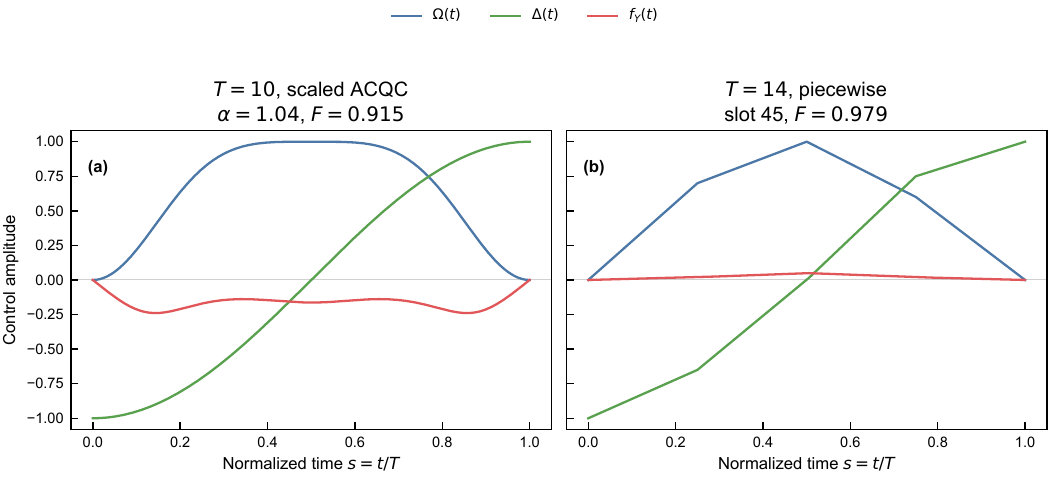}
  \caption{\textbf{Selected controls for the C6 per-time Rydberg search.} Two representative per-time winners are shown as functions of normalized time $t/T$: the $T=10$ selection uses a scaled ACQC-like $Y$ waveform on the smooth $\Omega(t)$ and $\Delta(t)$ schedules, while the $T=14$ selection uses piecewise-linear $\Omega(t)$ and $\Delta(t)$ schedules together with a multi-knot global-$Y$ pulse. These examples illustrate the two main control motifs selected across the seven total evolution times.}
  \label{fig:app-c6-controls}
\end{figure}

\begin{figure}[t]
  \centering
  \includegraphics[width=0.90\textwidth]{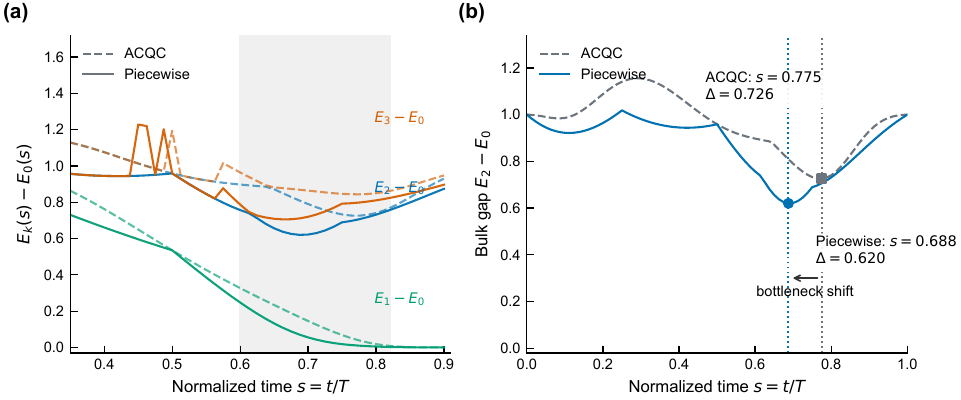}
  \caption{\textbf{C6 spectral diagnostic for the selected $T=14$ protocol.} \textbf{a}, Low-energy instantaneous spectrum for the ACQC baseline (dashed) and the selected piecewise schedule-plus-$Y$ protocol (solid), shown relative to the instantaneous ground energy. \textbf{b}, Bulk gap $E_2-E_0$ for the same C6, $T=14$ comparison. The selected piecewise protocol shifts the bulk-gap bottleneck from $s=0.775$ to $s=0.688$; the smaller gap value indicates that the improvement should not be interpreted as simple gap maximization.}
  \label{fig:app-c6-spectrum}
\end{figure}

The selected fixed-$T=14$ C6 protocol also provides a compact spectral diagnostic of the piecewise schedule-plus-$Y$ control relative to the ACQC baseline (Fig.~\ref{fig:app-c6-spectrum}).

The per-$T$ winner breakdown is as follows: at $T=8$ and $T=10$, the selected protocol is a scaled ACQC-like pulse with $\alpha=1.04$; at $T=12$, a weaker scaled-ACQC pulse ($\alpha<1$) is selected; at $T=14$, $T=16$ and $T=18$, the selected protocols are piecewise schedule deformations with global $Y$ pulses from the targeted or random piecewise families; and at $T=20$, a weaker ACQC scale $\alpha=0.52$ is again preferred. This behavior confirms that the preferred pulse-level correction depends on the available evolution time rather than being a single universal rescaling. The supplementary figure emphasizes that the per-time envelope is not produced by a single optimized waveform: different total times select different hardware-compatible control shapes.

\subsection{C10 supplemental global-$Y$ refinements}
\label{app:c10-details}

The C10 result in Fig.~\ref{fig:rydberg}b uses the fixed ten-site cycle instance and total time $T=30$. All candidates preserve the same Rydberg hardware restrictions as the main text: the interaction strength $V$ is fixed, the Rabi drive is non-negative and bounded by the smooth reference scale, and the only auxiliary control is a bounded global one-body quadrature $f_Y(t)\sum_iY_i$.

The explored candidate families do not represent an exhaustive global optimization over all possible control waveforms. Instead, they provide a structured, low-parameter space to systematically refine the initial C10 results. The smooth no-$Y$ candidate sets $f_Y(t)=0$ on the smooth $\Omega(t)$ and $\Delta(t)$ path. The symmetric five-knot family uses normalized time nodes $[0,0.25,0.5,0.75,1]$ and knot values $[0,0.4m,m,0.4m,0]$; its eleven points combine a coarse amplitude sweep with additional samples around the observed optimum, and include the prior $m=0.13$ pulse. The biased five-knot family keeps the same endpoints and midpoint but shifts weight earlier or later in the evolution. The seven-knot family uses $[0,1/6,1/3,1/2,2/3,5/6,1]$ with values $[0,0.2m,0.65m,m,0.65m,0.2m,0]$. The smooth beta-bump family replaces the piecewise-linear pulse by a differentiable endpoint-vanishing envelope $f_Y(t)=A B_{p,q}(t/T)$. Here $B_{p,q}(s)$ is the beta-distribution-shaped bump $s^p(1-s)^q/[s_*^p(1-s_*)^q]$, normalized to unit peak at $s_*=p/(p+q)$. Thus $A$ is the peak amplitude, $f_Y(0)=f_Y(T)=0$, and $p<q$ places the peak earlier than the midpoint; for the best C10 pulse with $p=1.6$ and $q=2.4$, the peak occurs at $s_*=0.4$.

\begin{figure}[t]
  \centering
  \includegraphics[width=0.95\textwidth]{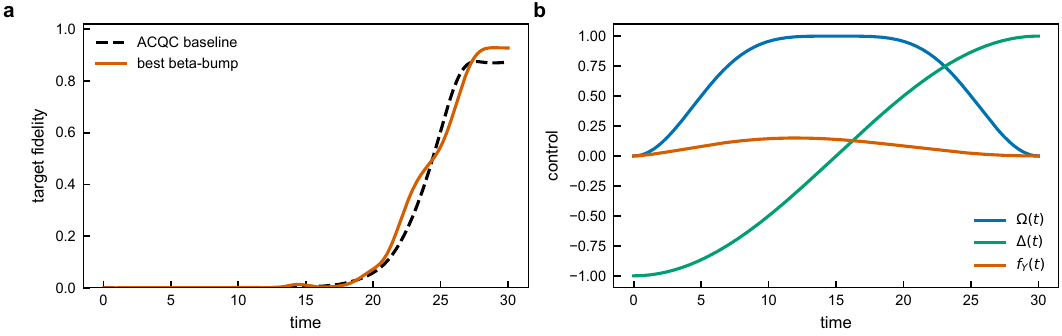}
  \caption{\textbf{Finite-time diagnostics for the C10 Rydberg protocols.} \textbf{a}, Target-ground-subspace fidelity versus time for the reproduced ACQC baseline and the best early-biased smooth beta-bump pulse. \textbf{b}, Control waveforms for the best protocol, which keeps the smooth $\Omega(t)$ and $\Delta(t)$ schedules fixed and uses the smooth beta-bump $Y$ envelope $f_Y(t)=AB_{p,q}(t/T)$ with $A=0.15$, $p=1.6$ and $q=2.4$, placing the peak at $s_*=p/(p+q)=0.4$. The final fidelity is $0.9275$, compared with $0.8725$ for ACQC.}
  \label{fig:app-c10-trajectory}
\end{figure}

Figure~\ref{fig:app-c10-trajectory} gives the corresponding finite-time diagnostics for C10, comparing the target fidelity and best-protocol control waveforms.

The ranking shows a clear waveform-design effect. The reproduced ACQC baseline gives final fidelity $0.8725$. The smooth no-$Y$ path already reaches $0.9066$. The prior symmetric five-knot $m=0.13$ pulse reaches $0.9165$. Increasing the symmetric five-knot amplitude improves this slightly near $m=0.15$--$0.17$, but overly strong pulses degrade performance. In this context, early-biased pulses place more $Y$-pulse weight before the midpoint of the evolution, whereas late-biased pulses place the corresponding weight after the midpoint. With this convention, early-biased pulses outperform late-biased pulses in the tested set. The best result comes from the smooth early-biased beta-bump envelope ($A=0.15$, $p=1.6$, $q=2.4$), reaching fidelity $0.9275$, outperforming the original analytic ACQC waveform.

\section{XXZ protocol formulae and schedule diagnostics}
\label{app:xxz-schedule}

\subsection{Catalyst, AH and commutator-CD Hamiltonians}
\label{app:xxz-formulae}

This section provides the theoretical details and extended results for Case 2 discussed in the main text, where the one-dimensional XXZ model is used as the target Hamiltonian. The baseline for all XXZ comparisons is the protocol reproduced from Ref.~\cite{xxz2026local}: the combined optimized-initialization, auxiliary-Hamiltonian and commutator-CD strategy, denoted OI+AH+CD in the text and ``Base'' in the XXZ plots. Throughout this section, the ``CD'' label denotes the scalar commutator-CD ansatz used in the simulations, not the exact many-body counterdiabatic Hamiltonian generated by the full adiabatic gauge potential. The refinement starts from this baseline, retains the OI and AH components, introduces an endpoint-vanishing target catalyst to the many-body interpolation path, and retunes the same commutator-CD coefficient. We denote the target Hamiltonian by
\begin{align}
H_f = H_{\rm XXZ}.
\end{align}
The optimized-initialization (OI) Hamiltonian is a separable local-field Hamiltonian,
\begin{align}
H_i^{\rm OI}
=-\epsilon\sum_{j=1}^{8}
\left[
\sin\theta_j^{\rm OI}\cos\phi_j^{\rm OI}\,X_j
+\sin\theta_j^{\rm OI}\sin\phi_j^{\rm OI}\,Y_j
+\cos\theta_j^{\rm OI}\,Z_j
\right],
\end{align}
where $\epsilon$ sets the local-field scale relative to the XXZ normalization in Eq.~(5), and is fixed to $\epsilon=1$ in the
  reproduced baseline; the angles $(\theta_j^{\rm OI},\phi_j^{\rm OI})$ define the optimized product-state initialization. The auxiliary-Hamiltonian (AH) term used with the OI path is a site-dependent longitudinal Zeeman field,
\begin{align}
H_{\rm aux}^{\rm OI}=\sum_{j=1}^{8}a_j^{\rm OI}Z_j,
\end{align}
with coefficients $a_j^{\rm OI}$ optimized in the baseline. The global sign in $H_i^{\rm OI}$ is chosen such that its ground state corresponds to the optimized product state; it is equivalent to the orientation convention of Ref.~\cite{xxz2026local} after redefining the local unit vectors. Thus both the initial Hamiltonian and the AH field are local one-body terms.

To incorporate the endpoint-vanishing target catalyst $H_f$ introduced in this work, we define a generalized controlled Hamiltonian path $H_{\rm cat}(s)$ with catalyst strength $\eta$:
\begin{align}
H_{\rm total}(t)={}&H_{\rm cat}(s)+H_{\rm cCD}(t),
\nonumber\\
H_{\rm cat}(s)={}&[1-\lambda(s)]H_i^{\rm OI}+\lambda(s)H_f
\nonumber\\
&+\lambda(s)[1-\lambda(s)]
\left(H_{\rm aux}^{\rm OI}+\eta H_f\right),
\nonumber\\
H_{\rm cCD}(t)={}&i\dot{\lambda}(t)\alpha
\left[H_{\rm cat}(s),\partial_{\lambda}H_{\rm cat}(s)\right],
\label{eq:catalyst}
\end{align}
where $s=t/T$ and
\begin{align}
\partial_{\lambda}H_{\rm cat}
=H_f-H_i^{\rm OI}+[1-2\lambda(s)]
\left(H_{\rm aux}^{\rm OI}+\eta H_f\right).
\end{align}
This is a first-commutator approximate CD correction with one fitted scalar $\alpha$; it is not a complete CD construction. The reproduced literature baseline (OI+AH+CD) corresponds exactly to the special case where the catalyst is absent ($\eta=0$) and the CD coefficient is set to its baseline optimized value ($\alpha=\alpha_{\rm base}$).

The parameters for the catalyst-enhanced protocols were selected by parameter grid scans over $\eta\in\{2,2.5,3,3.5,4,5,6\}$ and $\alpha\in\{-0.3,-0.2,-0.1,-0.07,-0.05,-0.03,0,0.05\}$, followed by local refinement. The final chosen parameters are $\eta=3.0$, $\alpha=-0.07$ for the $\Delta=1$ case and $\eta=3.0$, $\alpha=-0.08$ for the $\Delta=0.5$ case.

The subsequent schedule optimization preserves the catalyst and commutator-CD structure, replacing the reference smooth schedule $\lambda_{\rm ref}(s)$ with a mixed-power schedule,
\begin{align}
\lambda_{\rm mix}(s)={}&(1-q)\lambda_{\rm ref}(s)
\nonumber\\
&+q\frac{s^p}{s^p+(1-s)^p},
\nonumber\\
\lambda_{\rm ref}(s)={}&\sin^2\!\left[\frac{\pi}{2}\sin^2\!\left(\frac{\pi s}{2}\right)\right].
\label{eq:mixedschedule}
\end{align}
For the $\Delta=1$ result in Fig.~\ref{fig:xxz}b, the displayed Mix candidate uses $p=1.4$ and $q=0.5$; the Pow candidate uses the pure power component alone ($q=1$, $p=1.2$). For the $\Delta=0.5$ result in Fig.~\ref{fig:xxz}a, the Mix candidate uses $p=0.7$ and $q=0.25$, while the Pow candidate again uses the pure power component alone ($q=1$, $p=1.6$). To ensure strict boundary compliance for schedule deformations whose derivatives do not naturally vanish at the endpoints, appropriate windowing functions were applied. The Workbench proposed an endpoint-windowed commutator-CD variant $H_{\rm cCD}^{\rm win}(t)=\sin^2(\pi s)H_{\rm cCD}(t)$, which ensures the auxiliary correction vanishes at $s=0$ and $s=1$. This window was applied to the Sine candidate in Fig.~\ref{fig:xxz}b ($a=0.04$), where the sine deformation $\lambda_{\rm sine}(s)=\lambda_{\rm ref}(s)+a\sin(\pi s)$ does not preserve endpoint derivatives.

\subsection{Schedule shapes and instantaneous-overlap diagnostic}
\label{app:xxz-scheddiag}

\begin{figure}[t]
  \centering
  \includegraphics[width=0.72\textwidth]{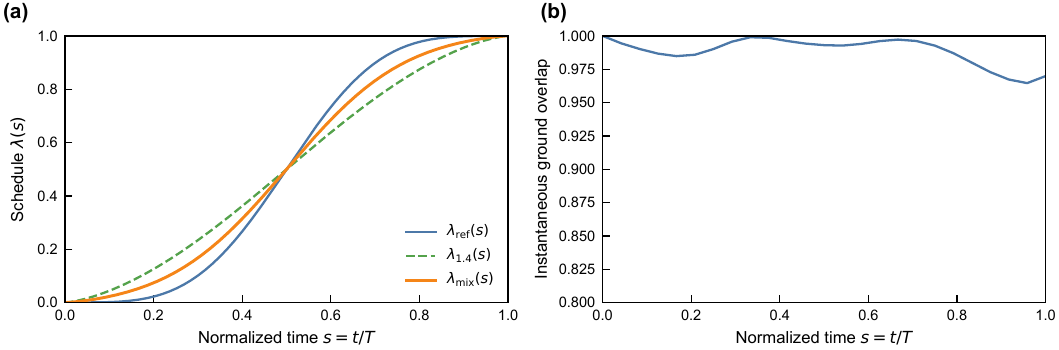}
  \caption{\textbf{Schedule deformation and instantaneous-overlap diagnostic for the $\Delta=1$, $T=3$ XXZ protocol.} \textbf{a}, The reference schedule $\lambda_{\rm ref}(s)$ is the smooth schedule inherited from the reproduced baseline. The power component with $p=1.4$ and the mixed schedule $\lambda_{\rm mix}(s)=(1-q)\lambda_{\rm ref}(s)+q\lambda_p(s)$ with $q=0.5$ keep the same endpoints and endpoint-vanishing derivatives while redistributing the rate at which the interpolation parameter advances through the path. \textbf{b}, Instantaneous ground-state overlap for the Fig.~\ref{fig:xxz}b mixed-schedule protocol, computed with respect to the instantaneous Hamiltonian including the catalyst and commutator-CD term.}
  \label{fig:app-xxz-schedule}
\end{figure}

Figure~\ref{fig:app-xxz-schedule} shows supplemental diagnostics for the endpoint-clean $\Delta=1$, $T=3$ XXZ mixed-schedule protocol displayed in Fig.~\ref{fig:xxz}b. Panel \textbf{a} plots the reference smooth schedule $\lambda_{\rm ref}(s)$, the power component with $p=1.4$ and their mixture $\lambda_{\rm mix}(s)$ with $q=0.5$ from Eq.~\eqref{eq:mixedschedule}. All three schedules preserve the same endpoints, $\lambda(0)=0$ and $\lambda(1)=1$; because $p>1$, the power component also has vanishing endpoint derivatives. Panel \textbf{b} reports the instantaneous ground-state overlap computed with respect to the instantaneous Hamiltonian of this same mixed-schedule protocol.

The XXZ optimization systematically evaluated several protocol families extending the reproduced baseline. As introduced above, SA denotes the simple adiabatic interpolation from the standard local initial Hamiltonian, OI denotes optimized initialization, AH denotes the endpoint-vanishing auxiliary Hamiltonian along the interpolation path, and CD denotes the approximate commutator-CD ansatz defined above. Among the baseline variants evaluated (SA, SA+AH, SA+CD, OI, OI+AH, OI+CD and OI+AH+CD), OI+AH+CD performed best and was selected as the reference baseline used as ``Base'' in the XXZ comparison plots. Building on this structure, subsequent experiments introduced the target catalyst (Cat+CD) and evaluated several nonlinear schedule deformations described in the last subsection.

\begin{figure}[t]
  \centering
  \includegraphics[width=0.82\textwidth]{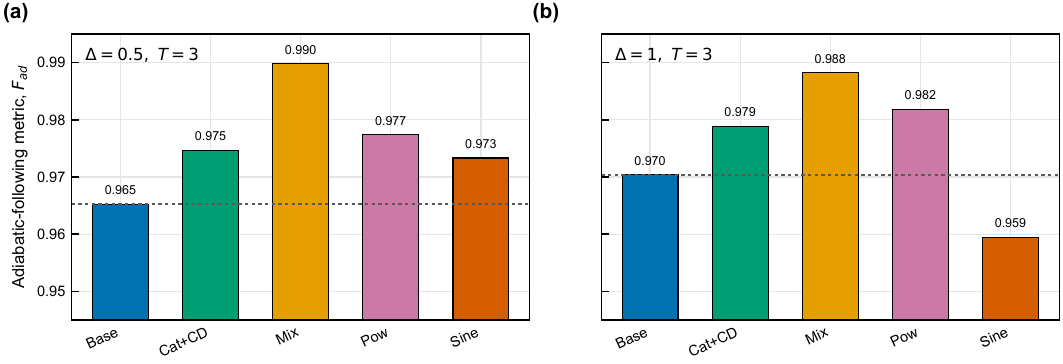}
  \caption{\textbf{Adiabatic-following metric for the discovered representative XXZ protocols.} \textbf{a}, $F_{ad}$ for the $\Delta=0.5$, $T=3$ protocols shown in Fig.~\ref{fig:xxz}a. \textbf{b}, $F_{ad}$ for the $\Delta=1$, $T=3$ endpoint-clean protocols shown in Fig.~\ref{fig:xxz}b. Dashed horizontal guide lines mark the reproduced OI+AH+CD baseline in each panel.}
  \label{fig:app-xxz-fad}
\end{figure}

Figure~\ref{fig:xxz} in the main text displays five representative endpoint-clean strategies for $\Delta=0.5, 1$ cases: the reproduced baseline (Base), the smooth schedule with target catalyst and CD (Cat+CD), and its combinations with mixed (Mix), pure power (Pow) and sine (Sine) schedules. The corresponding adiabatic-following metrics for these protocols are shown in Fig.~\ref{fig:app-xxz-fad}.

\section{TFIM instance details and neural generator results}
\label{app:tfim-architecture}

\subsection{Instance sampling and weighted-CD teacher}
\label{app:tfim-instances}

The TFIM instances use a rectangular open lattice. Random local fields $h_i$ and nearest-neighbour couplings $J_{ij}$ are independently sampled from a Gamma distribution with shape $4$ and scale $0.25$.  The rectangular open geometry is accessible for exact simulations at the sizes studied here, while still containing nonuniform local fields, spatially structured nearest-neighbour couplings and instance-dependent low-energy structure. 

For a Hamiltonian path $H(\lambda)$, exact counterdiabatic driving is defined by the adiabatic gauge potential, which is generally a nonlocal many-body operator. To ensure hardware compatibility, we restrict the auxiliary generator to a one-body ansatz, excluding independently tunable two-body or other nonlocal auxiliary operators. Furthermore, rather than minimizing the conventional variational action which weights all Hilbert-space transitions uniformly, we use the weighted variational CD framework~\cite{weightedcd2026local} to bias the correction toward the relevant low-energy subspace. 

Specifically, the target generator takes the one-body form
\begin{align}
V(\boldsymbol{\alpha}) = \sum_i \alpha_i(\lambda)Y_i,
\label{eq:onebodyY}
\end{align}
where the coefficients $\boldsymbol{\alpha}$ are obtained by minimizing the weighted variational action:
\begin{align}
\boldsymbol{\alpha}^{(K)}(\lambda)
={}&\arg\min_{\boldsymbol{\alpha}}
\left\|\partial_{\lambda}P_K[H(\lambda)]
- i\left[P_K[H(\lambda)],V(\boldsymbol{\alpha})\right]
\right\|_F^2,
\nonumber\\
P_K(H)={}&[H-E_K(\lambda)]^K.
\label{eq:weightedaction}
\end{align}
The conventional variational action corresponds to $K=1$; here we use $K=3$ to emphasize ground-state tracking. The energy shift $E_K(\lambda)$ was chosen above the instantaneous spectrum by grid search to concentrate the action on low-energy states while keeping the finite-dimensional least-squares problem stable. These coefficients $\alpha$ form the supervised labels used by the neural generators.

\subsection{Neural weighted-CD generator}
\label{app:tfim-neural-gen}

\begin{figure}[t]
  \centering
  \includegraphics[width=0.95\textwidth]{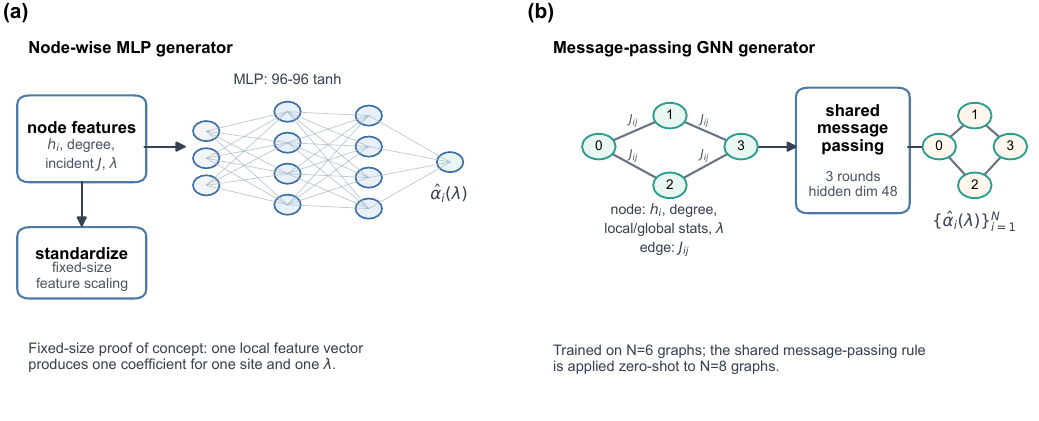}
  \caption{\textbf{Neural generators for weighted-CD coefficient paths.} \textbf{a}, The prototype uses a node-wise MLP regressor. For each node and each $\lambda$ value, local features are mapped to a single coefficient. \textbf{b}, The transfer model uses a message-passing GNN. Node features, edge features $J_{ij}$ and the coordinate $\lambda$ are processed by shared message-passing layers, producing node coefficients for graphs of different sizes. }
  \label{fig:app-tfim-architecture}
\end{figure}

The neural generator maps a TFIM instance and interpolation coordinate $\lambda$ to the one-body weighted-CD coefficient vector in Eq.~\eqref{eq:onebodyY}. The fixed-size prototype treats each site coefficient as a node-wise regression target and uses standardized graph-conditioned local features as input to a multilayer perceptron (MLP). The cross-size generator replaces this hand-crafted MLP by a message-passing graph neural network (GNN) implemented in JAX. The GNN uses node features containing local fields, degree, incident-coupling summaries, global field/coupling statistics and $\lambda$ features, together with edge features $J_{ij}$. Node features and edge couplings are updated through three shared message-passing steps with hidden dimension 48, allowing the same learned rule to be applied to larger graphs without changing the output head or trainable weights. Figure~\ref{fig:app-tfim-architecture} summarizes the MLP and GNN generator structures.

For both models, teacher labels were computed on 21 equally spaced $\lambda$ values. The MLP was trained on 48 random $2\times3$ ($N=6$) TFIM instances, validated on 12 additional $N=6$ instances and tested on 12 unseen $N=6$ instances. The GNN was trained on 24 random $N=6$ TFIM instances and validated on 6 additional $N=6$ instances; the zero-shot test used 6 random $2\times4$ ($N=8$) instances. At inference time, the learned coefficient path was evaluated on the $\lambda$ grid and linearly interpolated during time evolution.

\subsection{Quantitative results for neural generators}
\label{app:tfim-results}

Quantitative performance results are as follows.

\textbf{Fixed-size MLP ($N=6\to N=6$, $T=1.5$).} The MLP was trained on 48 random $N=6$ TFIM instances, validated on 12 additional $N=6$ instances and tested on 12 unseen $N=6$ test instances. On the test set, the MLP achieved coefficient prediction $R^2=0.989$ against the weighted-CD teacher labels. In the subsequent time-evolution simulation, the mean final ground-state fidelity was: no-CD baseline $0.147$, exact weighted-CD teacher $0.737$, MLP generator $0.732$. The MLP thus nearly matches the teacher fidelity distribution on held-out instances without reoptimization.

\textbf{Cross-size GNN ($N=6\to N=8$, $T=2.5$).} The GNN was trained on 24 random $N=6$ TFIM instances and validated on 6 additional $N=6$ instances. Zero-shot transfer was evaluated on 6 random $N=8$ test instances (no $N=8$ data seen during training). On these six unseen $N=8$ instances, the mean final ground-state fidelity was: no-CD baseline $0.197$, exact $N=8$ weighted-CD teacher $0.654$, $N=6$-trained GNN generator $0.680$. The final-fidelity advantage of the GNN over the teacher on this finite test set should not be interpreted as evidence that the GNN has learned a superior control path; it reflects the fact that the weighted variational action teacher is not identical to the final-time fidelity objective, and that a smooth graph-conditioned approximation can preserve or marginally improve the finite-time evolution in this regime. Coefficient prediction quality for the same $N=6$-trained GNN, evaluated across $N=6$ training, $N=6$ validation and zero-shot $N=8$ test splits, is summarized in Fig.~\ref{fig:tfim}c. The $N=8$ zero-shot test achieves $R^2=0.961$, confirming that the learned coefficient paths remain accurate on the larger lattice without retraining.

\section{Kick-start prompt for new Workbench tasks}
\label{app:prompt-assembly}

QOC-Workbench is intended to be reused on quantum optimal control settings beyond the three case studies reported here. A new user does not need to reproduce the full internal context of the original searches by hand. Instead, after cloning the repository, the user can provide a compact task description and ask the agent to follow the repository-local workflow documents and skills. The following example template is a suggested kick-start prompt for launching a new quantum optimization project.

\begin{verbatim}
I want to use this QOC Workbench to optimize a new quantum-control protocol.

Please follow the repo-local workflow docs and skills, especially:
- skills/new-hamiltonian-onboarding/SKILL.md
- skills/baseline-evaluation/SKILL.md
- skills/cross-paradigm-strategy-search/SKILL.md
- docs/new_task_backend_template.md
- docs/workbench_skills.md

Problem:
[Describe the physical problem and target task.]

Target Hamiltonian:
[Write H_target explicitly, including system size, boundary conditions,
parameters, and basis.]

Initial Hamiltonian or initial state:
[Write H_initial or say if you want the agent to choose a conservative
baseline and document it.]

Native controls and extensible search space:
[List experimentally native time-dependent control terms/channels,
their known amplitude bounds, smoothness/bandwidth limits, and hardware
restrictions. Treat these as the conservative starting control language,
not as a closed list. The agent may propose new auxiliary controls,
CD-inspired terms, catalysts, schedule transformations, composite
protocols, or amortized/neural generators, provided that each proposed
extension is explicitly defined, physically justified, checked against
the declared hardware and forbidden-control constraints, and evaluated
only through the protected simulator.]

Forbidden controls:
[List terms or shortcuts that must not be used.]

Exploration principle:
[Encourage cross-paradigm search: the agent should not merely tune the
listed native controls, but may enlarge or recombine the control
language when the proposal remains physically admissible and auditable.]

Metric:
[Define the primary objective, e.g. final ground-state fidelity, final
energy, approximation ratio, adiabatic-following fidelity. If unsure,
propose a metric and justify it before search.]

Baseline:
[Provide an existing baseline protocol/result if available. If none
exists, first create and evaluate a simple baseline.]

Goal:
[State the improvement target, e.g. beat baseline by 5%, reach fidelity
>= 0.99, or explore candidate families within a fixed budget.]

Budget:
[State simulation budget, runtime constraints, and system-size limits.]

Please first formalize the task spec and backend assumptions, run a
baseline artifact, then search for improved protocols only within the
declared constraints and auditable extension rules. Store official runs
under artifacts/ with summaries, diagnostics, candidate payloads, and
code snapshot for auditability.
\end{verbatim}

The detailed operational rules are intentionally kept in versioned repository files rather than repeated in the prompt. This keeps the user-facing request concise while allowing the Workbench to enforce a consistent procedure for task formalization, baseline evaluation, protocol search and artifact generation. Users can adapt the fields above to their Hamiltonian and hardware constraints while relying on the workflow documents to define the expected file layout, validation steps and audit trail.

\clearpage
\twocolumngrid

\end{document}